\documentclass[journal]{IEEEtran}
\usepackage{graphicx}
\usepackage{amsmath}
\usepackage{subcaption}
\usepackage{textcomp}
\usepackage{adjustbox}
\usepackage{tabularx}
\usepackage{color,soul}
\usepackage{enumitem}
\usepackage{float}
\usepackage[square,numbers]{natbib}
\newcommand{\etal}{\textit{et al. }}

\begin{document}

\title{Pericoronary adipose tissue attenuation as a predictor of functional severity of coronary stenosis}

\author{Marta~Pillitteri*, Guido~Nannini*, Simone~Saitta, Luca~Mariani, Riccardo~Maragna, Andrea~Baggiano, Gianluca~Pontone\dag, Alberto~Redaelli\dag%
    \thanks{Manuscript received February 19, 2025. The authors marked with * and \dag\ contributed equally to this work. The corresponding author is Guido Nannini (email: guido.nannini@polimi.it).}
    \thanks{Marta Pillitteri, Guido Nannini, Luca Mariani, and Alberto Redaelli are with Politecnico di Milano, Milan, Italy.}
    \thanks{Simone Saitta is with Department of Biomedical Engineering and Physics, Amsterdam UMC, Amsterdam, The Netherlands, and with University of Amsterdam, Amsterdam, The Netherlands.}
    \thanks{Riccardo Maragna, Andrea Baggiano, and Gianluca Pontone are with Centro Cardiologico Monzino IRCSS, Milan, Italy, and with University of Milan, Milan, Italy.}
}

\markboth{IEEE Transactions on Biomedical Engineering, Vol. XX, no. XX, MONTH YEAR}%
{Pillitteri \MakeLowercase{\textit{et al.}}: Pericoronary Adipose Tissue Attenuation}

\maketitle

\begin{abstract}
\textit{Objective:} This study aims to evaluate the functional significance of coronary stenosis by analyzing low-level radiomic features of the pericoronary adipose tissue (PCAT) surrounding the lesions, which are indicative of its inflammation status. 
\textit{Methods:} A dataset of 72 patients who underwent coronary computed tomography angiography (CCTA) was analyzed, with 3D segmentation and computational fluid dynamics (CFD) simulations from a prior study. Centerlines of the main epicardial branches were automatically extracted, and lesions identified using Gaussian kernel regression to estimate healthy branch caliber. PCAT features were computed per vessel following guideline recommendations and per lesion within a region extending radially for two vessel radii. Features like fat volume and mean attenuation (FAI) were analyzed for their relationship with CFD-derived hemodynamic biomarkers, such as fractional flow reserve (FFR) and wall shear stress (WSS). These features also informed a machine learning (ML) model for classifying potentially ischemic lesions.
\textit{Results:} PCAT exhibited, on average, higher attenuation in the presence of hemodynamically significant lesions (i.e., $\mathrm{FFR}<0.80$), although this difference was of limited statistical significance. The ML classifier, trained on PCAT features, successfully distinguished potentially ischemic lesions, yielding average accuracy of 0.84.
\textit{Conclusion:} PCAT attenuation is correlated with the functional status of coronary stenosis and can be used to inform ML models for predicting potential ischemia.
\textit{Significance:} PCAT features, readily available from CCTA, can be used to predict the hemodynamic characteristics of a lesion without the need for an invasive FFR examination.
\end{abstract}


\IEEEpeerreviewmaketitle

\section{Introduction}
\IEEEPARstart{C}{oronary} artery disease (CAD) is currently the leading cause of death worldwide \cite{okrainec2004coronary}. It consists of the occlusion of coronary vessels due to the buildup of fibro-lipidic plaque within the arterial walls, reducing the vessel's luminal area and causing diminished oxygen delivery to the myocardium, which can result in ischemia and, potentially, death. 
Functional assessment of coronary stenosis, using parameters such as fractional flow reserve (FFR)—defined as the ratio between the downstream stenotic pressure and the aortic pressure—is the gold standard in clinical practice for determining the severity of a lesion and making clinical decisions regarding the most appropriate treatment for the patient \cite{pijls2013fractional}. FFR examination is performed under X-ray guidance by inserting a pressure wire into a peripheral artery and guiding it to the lesion site in the coronary tree \cite{pijls1996measurement}. Despite its proven robustness, the widespread adoption of FFR is still limited due to the invasiveness and the cost of the procedure, thus making physician often rely on few indices directly quantifiable from medical imaging (e.g., stenosis grade). 
This prompted the development of alternative techniques for FFR examination, leveraging either numerical modeling \cite{taylor2013computational, sankaran2012patient} or, more recently, machine learning approaches \cite{suk2024deep, nannini2025learning}. \textit{In-silico} modeling leverages computational fluid dynamics (CFD) to simulate coronary flow within 3D anatomical coronary models reconstructed from coronary computed tomography angiography (CCTA) scans. This approach provides comprehensive, pointwise knowledge of velocity and pressure fields, enabling a virtual estimation of FFR, known as vFFR \cite{zarins2013computed, taylor2013computational}. Virtual FFR examination is a powerful non-invasive tool that is gaining traction in clinical practice for the assessment of CAD in stable patients, providing a robust alternative to invasive FFR measurement.
Moreover, CFD analysis allows the computation other functional parameters related to flow characteristics, which cannot be quantified with \textit{in-vivo} measurements, in a non-invasive fashion, such as wall shear stress (WSS), which is the tangential mechanical stress acting on the endothelial intima cells of the artery, and the trans-lesional FFR ($\Delta$FFR), i.e., the FFR drop across a stenosis, which are both associated with ischemic lesions requiring surgery, when exceeding critical cut-off values \cite{giannopoulos2018building}.
Yet, virtual FFR examination is still limited, mainly because of the cost of the service, offered by few commercial providers. Thus, identifying additional biomarkers capable of determining the functional status of a stenosis is a subject of significant scientific interest.

Pericoronary adipose tissue (PCAT) inflammation, which is evaluated through the fat attenuation index (FAI), has recently been suggested to be associated with the functional status of coronary stenosis \cite{napoli2023epicardial}. PCAT consists of the fatty tissue that surrounds coronary arteries and lies on the myocardial wall. In CCTA, fat is usually attenuated between -190 and -30 HU. An average attenuation of PCAT above -70 HU indicates an inflammatory state \cite{oikonomou2018non}. While it is well established that higher values of FAI on the main coronary branches-i.e., the left anterior descending (LAD), left circumflex (LCx) and right cornary artery (RCA)-are associated with an increased risk of cardiac mortality, and are frequently observed around ischemic lesions in patients with CAD \cite{napoli2023epicardial}, its relationship with the hemodynamic significance of the stenosis has not been clearly revealed.
The relationship between lesion significance and FAI has been investigated in various studies, without reaching a general consensus. Balcer \etal \cite{balcer2018pericoronary} found no association between FAI and lesion significance, while PCAT volume was strongly related to ischemia. Similarly, Yu \etal \cite{yu2023radiomics} did not find a relevant association between vFFR and FAI alone, but when leveraging models based on the analysis of radiomics features, the diagnostic performance was comparable to vFFR. Conversely, other authors suggested that increased values of FAI are associated with ischemic lesions (i.e., $\mathrm{FFR\leq0.8}$): Hoshino \etal \cite{hoshino2020peri} observed a significant difference in FAI values between lesions with $\mathrm{FFR\leq0.8}$ and $\mathrm{FFR>0.8}$, with higher FAI associated with potentially ischemic lesions. Similarly, Yan \etal \cite{yan2022pericoronary} and Yu \etal \cite{yu2020diagnostic} found that perivascular FAI is significantly higher in flow-limiting lesions causing ischemia.

While FAI and FFR are strong independent indicators of major adverse cardiovascular events, the nature of the FAI-FFR relationship remains uncertain. Thus, the primary objective of the present study is to investigate the relationship between FAI and the functional significance of coronary stenosis, as determined by vFFR, WSS and $\Delta \mathrm{FFR}$. To do this, we leveraged a dataset of CFD simulations performed on coronary artery anatomies and we extracted PCAT features from the proximal portions of each branch, as indicated by consensus guidelines \cite{ma2023evaluation}. Next, we conducted a per-lesion analysis, extracting for each stenosis morphological, hemodynamic, and PCAT features and assessing the correlation between PCAT inflammation and vFFR, WSS and $\Delta\mathrm{FFR}$. Finally, we used the PCAT features extracted to train a machine learning (ML) classifier to distinguish potentially ischemic lesions. 

In summary, the main contributions of this work are the following: \textit{i}) We present a fully automated and robust framework for quantifying PCAT features from CCTA; \textit{ii}) We investigate the association between PCAT features and functionally significant lesions, expanding the analysis to other functional parameters beyond FFR, predictors of ischemia; \textit{iii}) We propose a machine learning model that, fed with PCAT features extracted through our pipeline, can predict the hemodynamic significance of a stenosis.
The ultimate goal of this study is to assess whether PCAT features of the lesion can reliably determine its functional significance in the absence of FFR information.


\section{Methods}

\subsection{Dataset}
The study cohort was retrospectively analyzed at Centro Cardiologico Monzino (Milan, Italy) and comprised 72 patients who underwent CCTA for suspected CAD, accordingly to the guidelines of the European Association of Cardiovascular Imaging (EACVI). Images were acquired using a GE Revolution CT scanner (GE Healthcare, Milwaukee, Wisconsin) with ECG-gating and contrast enhancement. All acquired images had dimensions of  512 × 512 × 256 voxels, with pixel spacing ranging from 0.365 × 0.365 mm² to 0.4 × 0.4 mm², and slice thickness varying between 0.4 mm and 0.65 mm. 
The clinical and demographic characteristics of the patient population are summarized in Table \ref{tab:patient_characteristics}.
For each patient included in the dataset, the coronary lumen mesh and the corresponding pressure, vFFR, and WSS fields, defined at each mesh node, were available from previous works by our group. Briefly, CCTA scans were segmented to reconstruct the anatomy of the coronary arteries using a pre-trained two-stage convolutional neural network. Then, steady patient-taylored CFD simulations were run to simulate coronary flow during hyperemia, as described in Refs. \cite{nannini2024fully, nannini2024automated}.

\begin{table}[ht]
\caption{Patient characteristics and clinical data. Continuous variables were reported as mean $\pm\, $ standard deviation, categorical variables were reported as numbers (percentages) of patients or vessels.}
\label{tab:patient_characteristics}
\centering
\begin{tabularx}{0.49\textwidth}{X >{\centering\arraybackslash}X}
\hline
\textbf{Characteristics} & \textbf{Value} \\
\hline
Number of patients       & 72                    \\
Number of lesions        & 459                   \\
Age, year                & 64.7 \textpm\  8.9      \\
Male/Female, n           & 55/17                 \\
BMI, kg/m\(^2\)          & 26.7 \textpm\ 4.9            \\
Rest HR, beats/min       & 68.3 \textpm\ 9.8            \\
Systolic Pressure        & 141.5 \textpm\ 14.3          \\
Dyastolic Pressure       & 78.8 \textpm\ 7.7            \\
\textit{Risk factors}    &                       \\
Diabetes, n (\%)         & 12 (16.7)             \\
Hypertension, n (\%)     & 44 (61.1)             \\
Hypercholesterolemia, n (\%) & 48 (66.7)         \\
Smoking, n (\%)          & 30 (41.7)             \\
\textit{Vessel, n}       &                       \\
RCA, n (\%)              & 31 (20.6)             \\
LAD, n (\%)              & 60 (39.7)             \\
LCx, n (\%)              & 60 (39.7)             \\
\hline
\end{tabularx}
\end{table}

\subsection{Centerlines classification}\label{sec:cent_class}
\begin{figure*}[ht]
    \centering    \includegraphics[width=\linewidth]{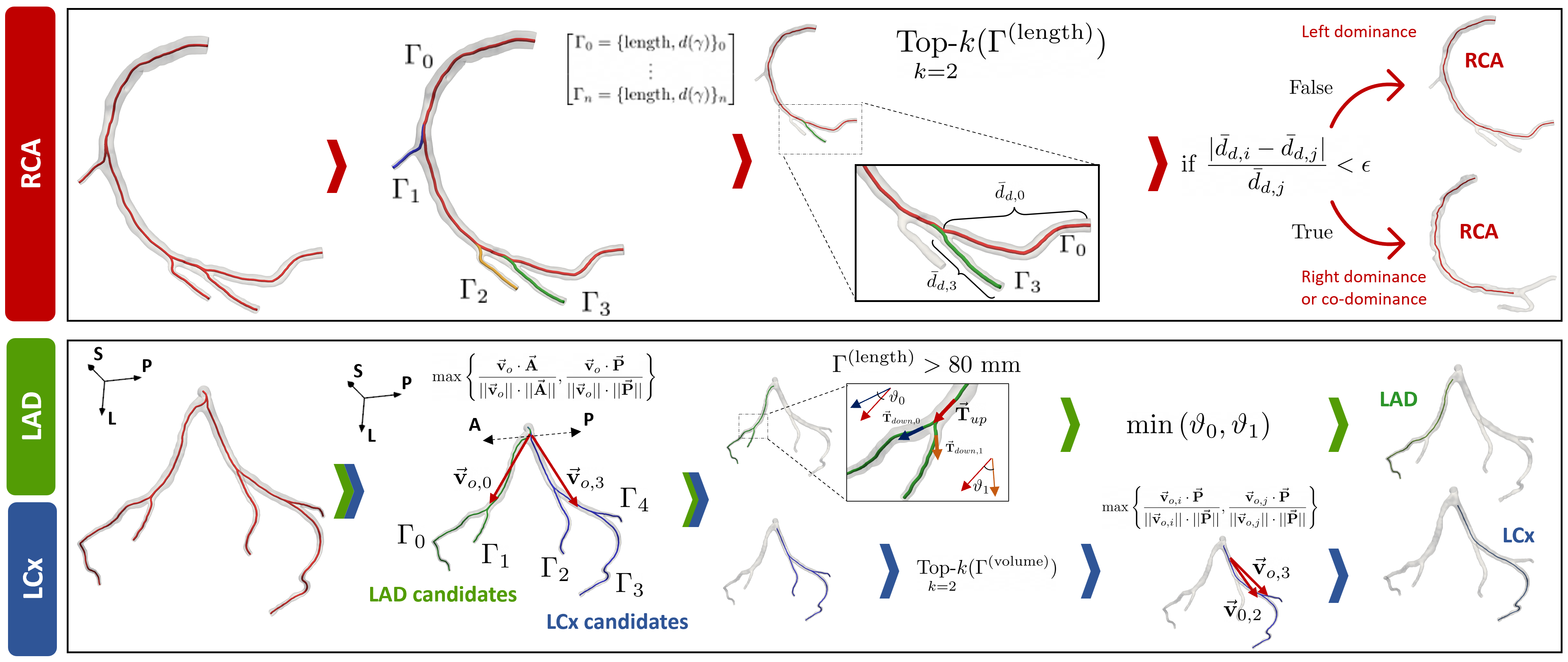}
    \caption{Schematic representation of algorithm described in Section \ref{sec:cent_class} for the automatic extraction of the centerlines of the three main coronary branches.}
    \label{fig:cline_classification}
\end{figure*}

Coronary centerlines were extracted using an in-house pipeline, detailed in Refs. \cite{nannini2024fully, saitta2022deep}. The pipeline automatically identifies seed nodes at the coronary ostium and target nodes at the end of each branch, and leverages the Vascular Modeling Toolkit (VMTK) library \cite{piccinelli2009framework} to extract the centerlines from the 3D mesh. 
Each centerline $\Gamma$ consist of an unstructured grid data with both global properties (e.g., the length) and local scalar and vector field, such as the diameter $d(\gamma)$ and the local tangent $\textbf{T}(\gamma)$, parameterized by the curvilinear abscissa $\gamma$.
Two groups of centerlines were extracted: one for the right coronary artery (RCA) and another for the left coronary artery (LCA). Each group includes all centerlines originating from the respective coronary ostium (right or left) and extending to each outlet, with coordinates expressed in the left-posterior-superior (LPS) reference system. A classification algorithm, schematized in Fig. \ref{fig:cline_classification}, was implemented to identify the centerlines corresponding to the three major epicardial coronary arteries: the right coronary artery (RCA), the left anterior descending artery (LAD), and the left circumflex artery (LCx). This algorithm was specifically designed for the available coronary lumen segmentations, which are truncated when the vessel caliber drops below 1.5 mm, a threshold that corresponds to the eligibility criterion for surgery \cite{sianos2005syntax}.

The identification of the RCA was based on the length of the centerline and mean caliber of the distal segment ($\bar{d_d}$). Based on these metrics, the following steps were conducted.
First, the two centerlines with the greatest lengths were considered for further analysis: these would typically correspond to either the posterior descending artery (PDA), the posterolateral branch (PLB), or in few cases, to the acute marginal branch (AMB). Then, the relative difference between the $\bar{d_d}$ of the two centerlines, measured from the bifurcation point, was computed as $\mathrm{Rel. \ Diff.}=\frac{|\bar{d}_{d,1} - \bar{d}_{d,2}|}{\bar{d}_{d,2}}$.

\begin{itemize}
    \item If $\mathrm{Rel. \ Diff.} > 40\%$ \cite{ballesteros2011right, villa_coronary_2016}, the smaller artery was designated as the AMB and the patient was considered with left coronary dominance. The largest artery was designated as the RCA.  
    \item Otherwise, the patient was considered at right coronary dominance or codominace, and the bifurcation point between the two centerlines (i.e., the PDA and PLB) was used as end point of the RCA. 
\end{itemize}

To classify the branches of the LCA tree, we first clusterd the centerline in LAD and LCX candidates, by computing the orientation of each centerline, defining a direction vector ($\vec{\textbf{v}}_o$). First, the bifurcation between the left anterior descending artery (LAD) and the left circumflex artery (LCx) was identified ($P_{\mathrm{bif}}$) as the first (i.e., the most proximal to the ostium) bifurcating node in the LCA tree. 
The left main branch was excluded from further analysis because its highly variable length across different patients makes it difficult to establish a standardized protocol for PCAT analysis \cite{zhang_peri-coronary_2024}.
We defined the direction of each branch as the vector originating from the bifurcation node and pointing to the center of mass ($P_{\mathrm{cm}}$) of the centerline: $\vec{\textbf{v}}_o=P_{\mathrm{cm}}-P_{\mathrm{bif}}$. Each centerline was classified as candidate LAD or LCX based on $\max{ \left\{ \frac{\vec{\textbf{v}}_o \cdot \vec{\textbf{A}}}{||\vec{\textbf{v}}_o||\cdot||\vec{\textbf{A}}||}, \frac{\vec{\textbf{v}}_o \cdot \vec{\textbf{P}}}{||\vec{\textbf{v}}_o||\cdot||\vec{\textbf{P}}||}\right\}}$, where $\vec{\textbf{A}}$ and $\vec{\textbf{P}}$ indicate the anterior and posterior direction, respectively. The argument of the maximum identify the direction of highest alignment of the vector $\vec{\textbf{v}}_o$. If $\vec{\textbf{v}}_o$ aligned the most with $\vec{\textbf{A}}$, the centerline was considered as a candidate LAD, otherwise (i.e., if $\vec{\textbf{v}}_o$ aligned with $\vec{\textbf{P}}$) the centerline was classified as candidate LCx.
\begin{itemize}
    \item For identifying the LAD, centerlines with a total length below 80 mm were excluded, as the typical lenght is 100 to 130 mm \cite{roberto_schetz_alves_origin_2022}. Among the remaining candidates the algorithm identifies the LAD centerline by first extracting all the bifurcation points and then analyzing local geometric properties at those nodes. For each candidate, the algorithm computes, at each bifurcation, the cosine similarity ($S_C$) by evaluating the dot product of the Frenet tangent vectors of one point upstream ($\vec{\textbf{T}}_{up}$) and one downstream ($\vec{\textbf{T}}_{down}$) the bifurcation, using a kernel of 1 cm, centered at the bifurcation node: $S_C= \frac{\vec{\textbf{T}}_{up} \cdot \vec{\textbf{T}}_{down}}{||\vec{\textbf{T}}_{up}||\cdot||\vec{\textbf{T}}_{down}||} $. This metric provides a measure of the alignment and smoothness of the centerline: for each bifurcation, the centerline with the highest cosine similarity (i.e., minimum angle) was iteratively selected, finally obtaining the LAD centerline.
    \item The algorithm identifies the LCx by filtering candidate centerlines based on their volume: the two candidates with the largest volumes (typically corresponding to the LCx, and either to one of the obtuse branch or the intermediate ramus, when present) were considered for further analysis. Between the two, the centerline with $\vec{\textbf{v}}_o$ more aligned to the $\vec{\textbf{P}}$ (i.e., with the highest dot product) was designated as LCx.
\end{itemize}
Following the automatic classification, a control step was included to verify the accuracy of the results, as the subsequent data processing relies heavily on this initial step.

\subsection{Stenosis identification}
The radii of the largest spheres inscribed within the lumen mesh were associated to each point of the selected centerlines. A robust weighted Gaussian kernel regression of the vessel radius was used to estimate the radius of an equivalent healthy vessel, using the methodology proposed by Shahzad \etal \cite{shahzad_automatic_2013}. 
For each point i of the vessel, the equivalent healthy radius $r_{i}^{h}$ is estimated as the weighted average of the observed radius $r_i$ with a Gaussian kernel with standard deviation $\sigma_x$ and the weights $w_i$:
\begin{equation}
r_{i}^{h} = \frac{\sum_{i'=1}^{n} \mathcal{N}(i' \mid i, \sigma_x) w_{i'} r_{i'}}{\sum_{i'=1}^{n} \mathcal{N}(i' \mid i, \sigma_x) w_{i'}}    
\label{eq:gaussreg}
\end{equation}

where $\mathcal{N}(i' \mid i, \sigma) = \frac{1}{\sigma \sqrt{2 \pi}} \exp \left( -\frac{(i' - i)^2}{2\sigma} \right)$.
Each weight is a measure of the distance between the effective radius function and a smoothed radius:

\begin{equation}
w_i = \mathcal{N}(r_i \mid r_{i}^{max}, \sigma_r)    
\end{equation}

$r_{i}^{max}$ is an average radius corrected by a constant $\kappa$ that is defined as the average of the maximum lesion prominence and the difference between the maximum and minimum radius across the vessel: 

\begin{equation}
    r_{i}^{max} = \frac{\sum_{i'=1}^{n} \mathcal{N}(i' \mid i, \sigma_{max}) r_{i'}}{\sum_{i'=1}^{n} \mathcal{N}(i' \mid i, \sigma_{max})} + \kappa.    
\label{eq:gauss_reg_k}
\end{equation}

The degree of stenosis ($SD$) was defined as the relative difference between the healthy radius and the stenotic radius:

\begin{equation}
    SD_i = 1 - \frac{r_i}{r_i^h}
\end{equation}

\begin{figure}
    \centering
    \includegraphics[width=\columnwidth]{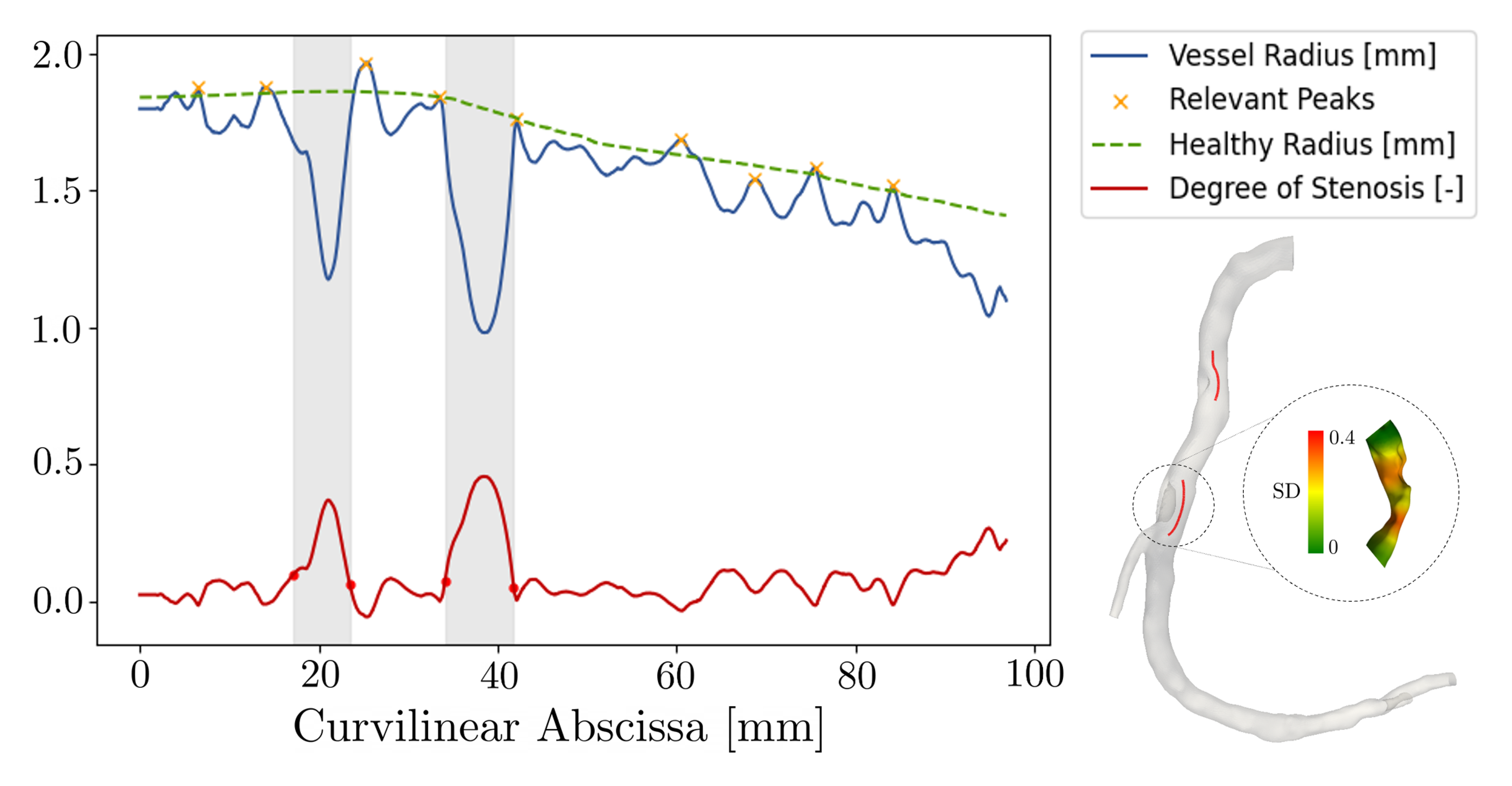}
    \caption{Example of stenosis detection in an RCA, using Gaussian regression to estimate the healthy radius (green line) from the vessel radius (blue). The red line represents the degree of stenosis. Grey bands identify the lesions.}
    \label{fig:stenosis_detection}
\end{figure}

The parameters $\sigma_x$, $\sigma_{max}$, $\sigma_r$ were optimized for each patient to minimize a specifically designed loss function. The optimization process was conducted in two stages using the SciPy library. First, we performed a grid-based brute-force search over predefined bounds to identify a preliminary set of optimal parameters. The brute-force method evaluated the mean squared error (MSE) between the estimated healthy radii and the observed radii at filtered peak locations, which were determined using distance-based peak detection criteria (i.e., setting a minimum distance between two peaks of 2.5 diameters). The filtered peaks included only the most prominent ones.
Second, the preliminary parameters obtained from the brute-force search were refined using the L-BFGS-B algorithm, a gradient-based optimization method. This step utilized the output from the brute-force search as the initial guess, allowing for finer parameter tuning within the specified bounds.
The same MSE was used as loss function to be minimized. This dual-stage approach ensured both robustness (via brute-force search) and precision (via gradient-based refinement) in parameter estimation.
The regression algorithm produced a function defined at each point along the centerline, representing the stenosis grade $SD(\gamma)$, as in Fig. \ref{fig:stenosis_detection}.

\begin{figure}
    \centering    \includegraphics[width=\columnwidth]{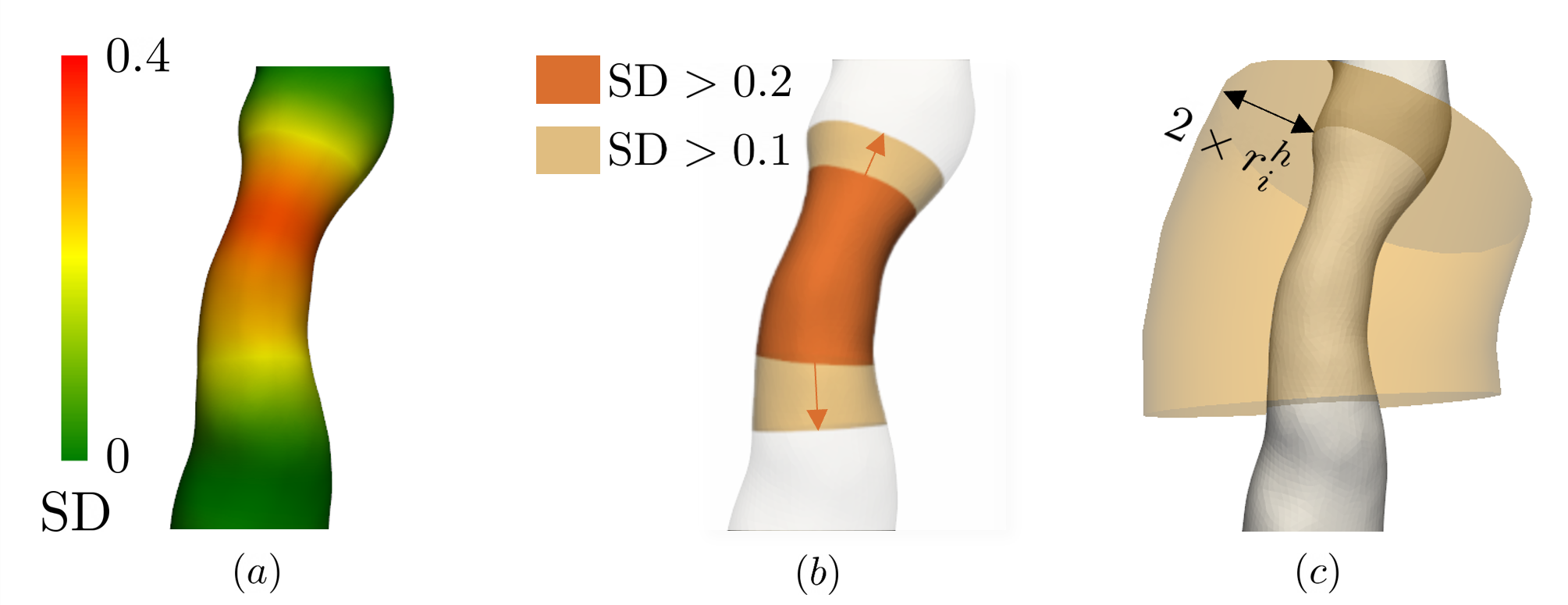}
    \caption{Definition of a stenotic region and corresponding ROI for PCAT per-lesion analysis. (a) Stenosis degree (SD) map visualized on an arterial segment; (b) Lesion detected as a portion with SD$>$0.2, extended up to SD$>$0.1; (c) ROI used for PCAT features analysis.}
    \label{fig:roiFAIstenosis}
\end{figure}

Stenotic lesions were identified detecting groups of points of the centerline for which $SD(\gamma)>20\%$. The lesion portion was then extended including adjacent points for which $SD(\gamma)>10\%$, similarly to Ref. \cite{fossan2018uncertainty} (Fig. \ref{fig:roiFAIstenosis}). The detected lesions were finally filtered, based on the following criteria:
\begin{enumerate}
    \item If the detected stenosis was within 2.5 diameter from the coronary ostium, it was excluded to account for boundary effect (i.e., the physiological decrease of the coronary caliber from the ostium region to the proper artery);
    \item Stenotic regions within 2.5 diameters of the branch endpoint were excluded, as the regression algorithm's accuracy deteriorated in the most distal branch segments due to boundary effects;
    \item To ensure clinical relevance, regions with a stenosis length shorter than 2 mm were excluded (in analogy to the SCCT recommendations regarding the minimum vessel diameter for quantitative plaque analysis \cite{nieman_standards_2024}).
\end{enumerate}

For each detected stenotic region, the following morphological parameters were calculated:
\begin{itemize}
    \item \textbf{Maximum stenosis degree}: The highest value of \(\text{SD}\) within the stenotic region.
    \item \textbf{Stenosis length}: The difference between the abscissa coordinates of the upstream and downstream lesion boundaries.
    \item \textbf{Minimum lumen area}: The area of the most stenotic section. 
    \item \textbf{Distance from ostium}: The distance of the most stenotic section from the right or left ostium. 
    \item \textbf{Tortuosity}: The ratio of the actual lesion length along the curvilinear abscissas and the straight-line distance between the lesion’s endpoints.
\end{itemize}

\subsection{Analysis of PCAT features}\label{sec:fai_feat}

\subsubsection{ROI definition}
\begin{figure*}[ht]
    \centering
    \includegraphics[width=\linewidth]{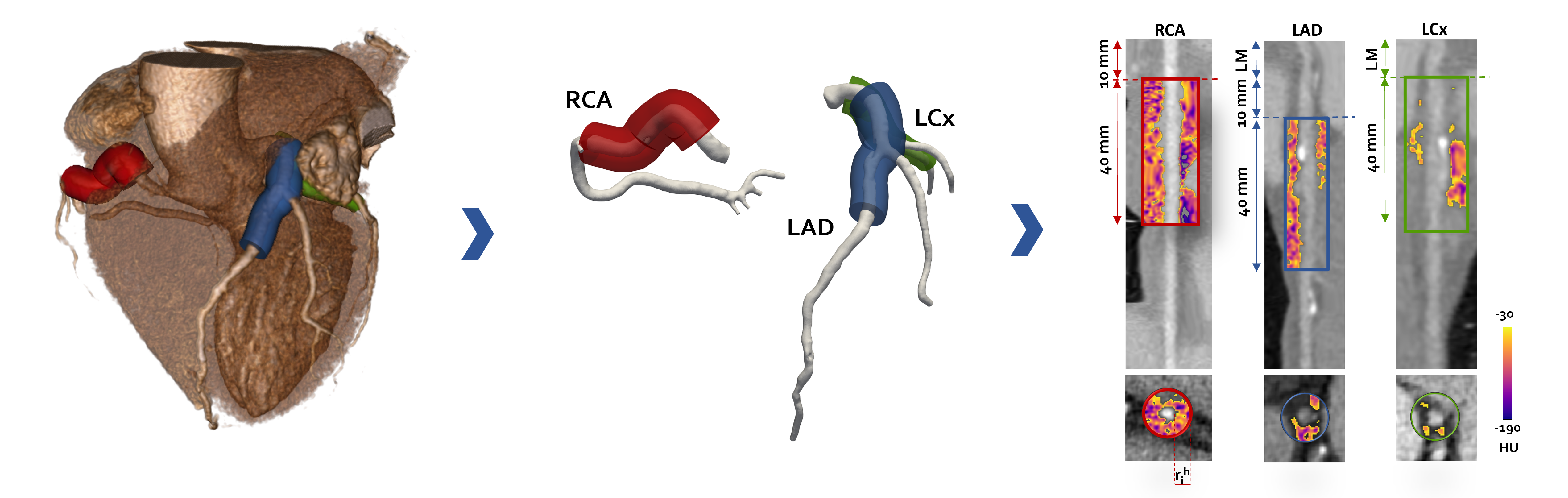}
    \caption{Per-vessel definition of region of interest (ROI) for the analysis of the pericoronary adipose tissue of the three main branches: RCA, LAD, LCx. Figure shows the ROIs in the whole cardiac anatomy and on the isolated segmentation of coronary arteries, and the multiplanar reformations of the CCTA for these ROIs.}
    \label{fig:ROIvessel}
\end{figure*}

The analysis of pericoronary adipose tissue (PCAT) was conducted on a per-vessel and per-lesion basis, employing standardized approaches to ensure reproducibility and accuracy. In each case, the region of interest (ROI) extended radially outward from the external wall of the vessel by a distance equivalent to the vessel diameter, reflecting the established anatomical boundaries of PCAT.
For vessel-based analysis (Fig. \ref{fig:ROIvessel}), the ROI was defined as a cylindrical volume with a fixed longitudinal length of 40 mm. The starting point varied based on the coronary artery being analyzed: for the right coronary artery (RCA), the ROI started 10 mm distal to the ostium; for the left anterior descending artery (LAD), it was located 10 mm distal to the bifurcation of the left main coronary artery (LMCA); and for the left circumflex artery (LCx), it started immediately distal to the LMCA bifurcation. These starting points were selected to avoid the effects of the aortic root. This methodology ensured consistent sampling across different vascular territories.

For lesion-based analysis (detected as in Fig. \ref{fig:stenosis_detection}), the ROI was delineated along the curvilinear abscissa corresponding to the extent of the stenosis, as defined in the preceding section.
The ROI was modeled as a polydata structure, representing a cylinder with a radius three times the estimated healthy vessel radius. The polydata structure was subsequently converted into a binary mask, which was rigidly registered to the CCTA imaging dataset. To isolate PCAT, the lumen voxels were subtracted, leaving only the voxels within the perivascular region.
To differentiate PCAT from other tissues and avoid contamination by non-adipose structures, a Hounsfield Unit (HU) threshold was applied. Voxels with attenuation values between -190 and -30 HU were retained, as this range is well-established in the literature to represent adipose tissue. 

\subsubsection{Features computation}
For each defined ROI, a comprehensive set of parameters related to PCAT is systematically analyzed, as these serve as critical biomarkers for assessing coronary lesion inflammation. The analysis includes the calculation of the mean attenuation of PCAT (FAI), the percentiles of the adipose tissue's Hounsfield Unit (HU) distribution (10, 25, 50, 75, 90, 95), the proportion of the ROI occupied by adipose tissue, and the volumetric measurement (in mm³) of the adipose tissue isolated within the defined ROI. 

\subsection{Statistical analysis}
Statistical analyses were conducted using Python SciPy 1.10.1 statistics library \cite{2020SciPy-NMeth}. The Shapiro-Wilk test was used to asses data normality. Continuous variables with normal distributions are reported as mean $\pm$ standard deviation; non-normally distributed data are reported as median and quartiles. Continuous hemodynamic variables were converted into categorical variables based on cut-off values identified in the literature, specifically from the EMERALD study, which established optimized thresholds for distinguishing between culprit and non-culprit lesions.
The Student's t-test was employed for data with a normal distribution, while the Mann–Whitney U-test was applied to non-normally distributed data.

\subsection{ML classificator training}\label{sec:ml_class}
The morphological and PCAT-related features extracted as described in Section \ref{sec:fai_feat} were used to train a ML classifier to identify potentially culprit lesions, based on the value of three functional parameters: vFFR, $\Delta\mathrm{FFR}$ and WSS, when informed with geometric and PCAT features. Different ML models adopted for analogous tasks \cite{corti2024enhancing} were tested to select the most suitable architecture (see Supplementary Materials), and a multi-layer perceptron (MLP) backend was ultimately selected. Each lesion was assigned a categorical value (i.e., 0 or 1) based on the cut-off threshold reported in the EMERALD trial \cite{giannopoulos2018building} for each specific variable (Table \ref{tab:cutoff_emerald}). Additionally we defined a fourth class that we called high risk stenosis (HRS), which we considered functionally relevant if at least 2 out of the 3 biomarkers were relevant.

The same MLP architecture, consisting of 4 layers with 128 channels, was trained separately on four distinct datasets. Three datasets included lesions in each of the main branches (LAD, LCx, and RCA), while the fourth dataset included lesions across all three branches. Each dataset was split into training, validation, and test sets using an 80/10/10 ratio, ensuring a balanced representation of functional and non-functional severe lesions. The models were trained for 300 epochs using the ADAM optimizer with an initial learning rate of 0.001 and exponential decay with $\gamma_{decay}=0.99$ to minimize the cross-entropy loss between predicted and target values. To enhance the robustness of training, feature selection was performed prior to each training session using a recursive elimination approach based on a logistic regression estimator to identify the 7 most representative features. Sensitivity analysis of the model dimensions and other hyperparameters is available in the Supplementary Materials. To evaluate the model's performance, the F1-score, accuracy, and area under the curve (AUC) were computed during inference.

\begin{table}[ht]
\caption{Criteria used to classify lesions between functionally severe and non, based on FFR, WSS, and $\Delta$FFR values.}
\centering
\resizebox{0.49\textwidth}{!}{
\begin{tabular}{c|cccc}
\hline
\textbf{Condition} & \textbf{FFR} & \textbf{WSS} & \textbf{$\Delta$FFR} & \textbf{HRS} \\
\hline
Severe (1) & $\leq$ 0.80 & $\geq$ 15.47 Pa & $\geq$ 0.06 & $\geq$ 2 out of 3\\
Non-severe (0) & $>$ 0.80 & $<$ 15.47 Pa & $<$ 0.06 & Otherwise\\
\hline
\end{tabular}}
\label{tab:cutoff_emerald}
\end{table}

\section{Results}

\subsection{Automatic Centerline Classification}
The algorithm correctly identified the RCA centerline in 32 out of 33 right coronary vessels, achieving an accuracy of 97\%. Additionally, it successfully distinguished cases of right dominance or codominance, where the RCA bifurcates into the PDA and PLB, by accurately identifying the RCA centerline upstream of the bifurcation. In left dominance cases, where the RCA exhibited no significant terminal bifurcations, the algorithm accurately identified the vessel along its entire epicardial extension.
The centerlines of the LCA were always correctly clustered into candidate LAD and candidate LCx categories, respectively including the LAD and all the diagonal branches, and the LCx and all the obtuse marginal branches. Among the candidates, LAD was correctly classified in 58 out of 60 cases, achieving an accuracy of 96.67\%, while LCx was correctly classified in 53 out of 60 cases, achieving an accuracy of 88.33\%.
After automatic segmentation, a control step was introduced to ensure the reliability of the subsequent analysis, allowing for manual correction of the classification.

\subsection{Stenosis identification}
Regression was independently performed on each anatomy, by optimizing the paramaters $\sigma_x$, $\sigma_{max}$, $\sigma_r$, which resulted (expressed in mm as mean [min, max]), respectively, 10.4 [10.0; 17.5], 21.5 [3.67; 50.0], 0.296 [0.250, 0.556].
A total of 457 stenoses were identified: 96 (21\%) in the RCA, 223 (48.8\%) in the LAD, and 138 (30.2\%) in the LCx. Table \ref{tab:stenosis_parameters} summerize the main geometric and PCAT features computed on each lesion, reported as mean [5$^{\text{th}}$ percentile; 95$^{\text{th}}$ percentile]. A comprehensive table including all extracted features is provided in the Supplementary Materials.

\begin{table}[ht]
\caption{Summary statistics for RCA, LAD and LCx morphometric parameters. SD = stenosis degree. MLA = Minimal luminal area. FAI = Fat attenuation index. PCAT = Pericoronary adipose tissue.}
\label{tab:stenosis_parameters}
\centering
\begin{tabularx}{0.49\textwidth}{X >{\centering\arraybackslash}X}
\hline
\multicolumn{2}{c}{\textbf{\textit{LAD}} (\textit{n}=223)} \\
\hline
Lesions per vessel (\textit{n}) & 3.66 $\pm$ 1.83\\
SD (\%) & 28.57 [20.70; 52.87] \\
MLA (mm\textsuperscript{2}) & 3.20 [1.19; 9.08] \\
Length (mm) & 6.19 [2.60; 14.77] \\
Distance from Ostium (mm) & 45.25 [16.54; 110.16] \\
Tortuosity & 0.97 [0.92; 0.99] \\
FAI (HU) & -82.0 [-98.0; -64.1] \\
PCAT volume (\%) & 51.67 [23.51; 72.93] \\
\hline
\multicolumn{2}{c}{\textbf{\textit{LCx}} (\textit{n}=138)} \\
\hline
Lesions per vessel (\textit{n}) & 2.38 $\pm$ 1.40\\
SD (\%) & 27.99 [20.69; 50.54] \\
MLA (mm\textsuperscript{2}) & 2.75 [1.00; 9.19] \\
Length (mm) & 6.48 [3.03; 13.94] \\
Distance from Ostium (mm) & 44.47 [17.31; 109.20] \\
Tortuosity & 0.97 [0.86; 0.99] \\
FAI (HU) & -75.0 [-95.5; -56.5] \\
PCAT volume (\%) & 30.76 [8.77; 50.73] \\
\hline
\multicolumn{2}{c}{\textbf{\textit{RCA}} (\textit{n}=96)} \\
\hline
Lesions per vessel (\textit{n}) & 3.43 $\pm$ 1.87\\
SD (\%) & 28.18 [20.64; 52.04] \\
MLA (mm\textsuperscript{2}) & 5.99 [2.43; 13.39] \\
Length (mm) & 7.17 [3.69; 16.43] \\
Distance from Ostium (mm) & 46.03 [8.24; 109.43] \\
Tortuosity & 0.97 [0.91; 0.99] \\
FAI (HU) & -77.8 [-93.3; -55.9] \\
PCAT volume (\%) & 49.9 [19.6; 81.2] \\
\hline
\end{tabularx}
\end{table}

\subsection{Association between PCAT features and invasive FFR measurements.}
Among the 151 vessels analyzed, 20.5\% of RCA, 24.1\% of LAD, and 23.1\% of LCx were classified as possible ischemia-causing vessels (FFR $\leq$ 0.80).

\subsubsection{Per-vessel analysis}
\begin{figure}
    \centering
    \begin{subfigure}{\columnwidth} 
        \centering
        \includegraphics[width=\columnwidth]{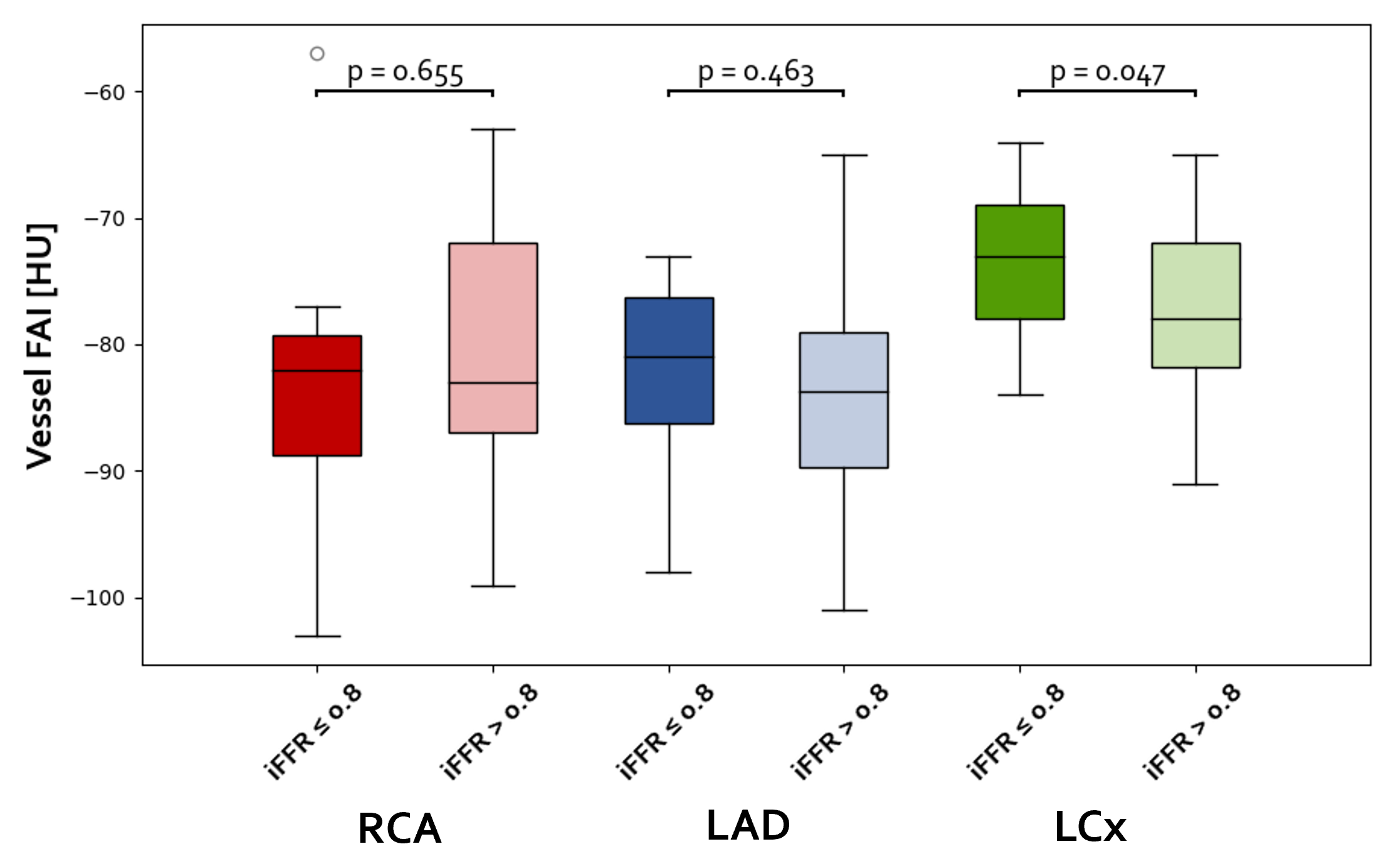}
        \caption{}
        \label{fig:VesselFAI-a}
    \end{subfigure}
    \vspace{0.5cm}
    \begin{subfigure}{\columnwidth} 
        \centering
        \includegraphics[width=\columnwidth]{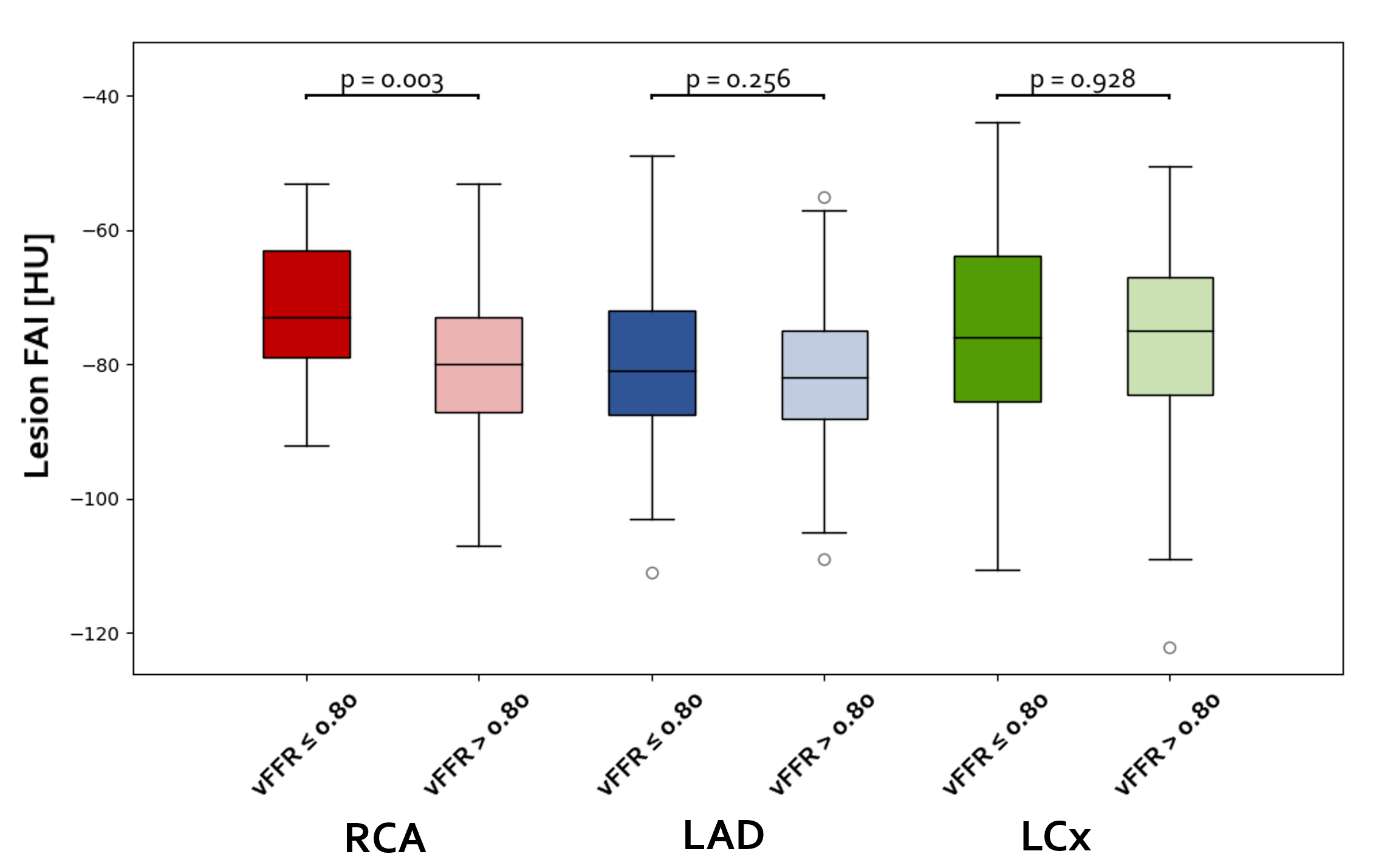}
        \caption{}
        \label{fig:VesselFAI-b}
    \end{subfigure}
    \caption{Fat attenuation index (FAI) in the main epicardial coronary vessels. (a) Per-vessel in ischemic and non-ischemic patient, classified based on invasive FFR. (b) Per-lesion in ischemic and non-ischemic patient, classified based on virtual FFR (from CFD).}
    \label{fig:VesselFAI}
\end{figure}

The relationship between per-vessel FAI values and FFR derived from ICA was investigated. No statistically significant differences in FAI values were observed between flow-limiting (FFR $\leq$ 0.80) and non-flow-limiting vessels for the RCA or LAD. However, a significant difference was identified for the LCx. For each of the three vessels, the median FAI values were higher in flow-limiting vessels (FFR $\leq$ 0.80), although the differences did not reach statistical significance [RCA: -83 vs. -82 (p = 0.655); LAD: -83.8 vs. -81 (p = 0.463); LCx: -78 vs. -73 (p = 0.047)] (Fig. \ref{fig:VesselFAI}). 

\subsubsection{Per-lesion analysis} At the lesion level, FAI values were significantly higher in lesions with hemodynamically significant stenosis (FFR $\leq$ 0.80) only in the RCA [RCA: -80 vs. -73 (p = 0.003)], while no significant differences were observed for the LAD [-82 vs. -81 (p = 0.256)] or LCx [-75 vs. -76 (p = 0.928)].
Regarding WSS values above or below the threshold of 15.47 Pa, no significant differences in FAI were found for the RCA (-77.00 vs. -79.50; p = 0.197) or LAD (-80.50 vs. -82.00; p = 0.176). Conversely, a significant difference was observed for the LCx (-77.00 vs. -72.50; p = 0.033).

For the Delta FFR analysis, although no statistically significant differences were observed for any of the vessels (RCA: p = 0.293, LAD: p = 0.361, LCx: p = 0.069), some trends were noted. In both the RCA and LAD, the median FAI values were higher in cases with Delta FFR above the threshold (RCA: -74.50 vs. -78.00; LAD: -80.00 vs. -82.00), suggesting a possible association between increased Delta FFR and higher FAI.

\subsection{Classification of culprit lesion based on PCAT features}

Table \ref{table:test_set_vessel} and \ref{table:test_set_all} report the test set result of the trained classifier.
When focusing on a specific branch, the trained model were able to achieve a good diagnostic accuracy identifying culprit lesions both on the basis of FFR and WSS, yielding accuracy of [0.78-0.93] and [0.80-0.87], respectively, with fair-to-high values of F1-score and AUC. When $\Delta\mathrm{FFR}$ was used as the index for classifying culprit lesions, the model's performance worsened, resulting in lower values across all metrics. In the classification of HRS the model yielded good results in the RCA, achieving an accuracy of 0.90, AUC=0.83 and F1-score=0.8.
Using lesion from any branch to train our classificator, the performance where in line with the results per-vessel: the diagnostic accuracy of the model was the best when using FFR and WSS as criteria for identifying severe lesions, yielding an accuracy of 0.79 and 0.87 respectively, with good AUC and F1-scores. Despite a good accuracy achieved to classify lesions based on the $\Delta\mathrm{FFR}$, low values of F1 and AUC were obtained, while culprit lesion based on HRS criterion were poorly classified.

\begin{table}[ht]
\caption{Performance metrics for the classification of functionally severe lesions in LAD, LCx and RCA, based on lesion geometric and PCAT features}
\centering
\resizebox{0.49\textwidth}{!}{
\begin{tabular}{l|cccc}
\hline
 \textbf{\textit{LAD}} & \textbf{FFR$\leq0.8$} & \textbf{WSS$\geq15.47$} & \textbf{$\Delta$FFR$\geq0.06$} & \textbf{HRS} \\
\hline
Test Loss $\downarrow$ & 0.42$\pm$0.46 & 0.48$\pm$0.52 & 0.51$\pm$0.75 & 0.56$\pm$0.59 \\
F1-Score $\uparrow$ & 0.71 & 0.90 & 0.60 & 0.63 \\
Accuracy $\uparrow$ & 0.78 & 0.87 & 0.83 & 0.74 \\
AUC $\uparrow$ & 0.78 & 0.83 & 0.71 & 0.71 \\
\hline
 \textbf{\textit{LCx}} & \textbf{FFR$\leq0.8$} & \textbf{WSS$\geq15.47$} & \textbf{$\Delta$FFR$\geq0.06$} & \textbf{HRS} \\
\hline
Test Loss $\downarrow$ & 0.15$\pm$0.18 & 0.43$\pm$0.23 & 0.66$\pm$0.74 & 0.56$\pm$0.57 \\
F1-Score $\uparrow$ & 0.80 & 0.88 & 0.50 & 0.40 \\
Accuracy $\uparrow$ & 0.93 & 0.86 & 0.57 & 0.57 \\
AUC $\uparrow$ & 0.83 & 0.85 & 0.63 & 0.53 \\
\hline
\textbf{\textit{RCA}} & \textbf{FFR$\leq0.8$} & \textbf{WSS$\geq15.47$} & \textbf{$\Delta$FFR$\geq0.06$} & \textbf{HRS} \\
\hline
Test Loss $\downarrow$ & 0.35$\pm$0.27 & 0.39$\pm$0.68 & 0.30$\pm$0.35 & 0.31$\pm$0.28 \\
F1-Score $\uparrow$ & 0.67 & 0.86 & 0.50 & 0.80 \\
Accuracy $\uparrow$ & 0.80 & 0.80 & 0.80 & 0.90 \\
AUC $\uparrow$ & 0.76 & 0.75 & 0.68 & 0.83 \\
\hline
\end{tabular}}
\label{table:test_set_vessel}
\end{table}

\begin{table}[ht]
\caption{Performance metrics for the classification of functionally severe lesions, including all branches in the dataset, based on lesion geometric and PCAT features}
\centering
\resizebox{0.49\textwidth}{!}{
\begin{tabular}{l|cccc}
\hline
\textbf{\textit{All}} & \textbf{FFR$\leq0.8$} & \textbf{WSS$\geq15.47$} & \textbf{$\Delta$FFR$\geq0.06$} & \textbf{HRS} \\
\hline
Test Loss $\downarrow$ & 0.38$\pm$0.78 & 0.54$\pm$0.92 & 0.40$\pm$0.41 & 0.55$\pm$0.46 \\
F1-Score $\uparrow$ & 0.79 & 0.93 & 0.46 & 0.41 \\
Accuracy $\uparrow$ & 0.79 & 0.87 & 0.83 & 0.68 \\
AUC $\uparrow$ & 0.85 & 0.75 & 0.72 & 0.62 \\
\hline
\end{tabular}}
\label{table:test_set_all}
\end{table}

\section{Discussion}
We presented a fully automated framework for the analysis of major coronary vessels, which includes vessel classification, identification of stenotic lesions, and quantification of morphometric, hemodynamic, and PCAT-related features. 
The coronary branch classification algorithm was developed to produce results consistent with the Society of Cardiovascular Computed Tomography (SCCT) coronary segmentation diagram \cite{leipsic_scct_2014}. Overall, a very good accuracy was achieved (RCA: 97\%, LAD: 96.67\%, LCx: 88.33\%). The algorithm reliably identifies the three main epicardial vessels, which are of paramount clinical significance in the study of coronary stenosis and the only relevant segments for examining pericoronary adipose tissue \cite{ma2023evaluation}. In the few cases of misclassification, the control step ensured that errors at this stage of the pipeline did not affect the overall analysis. The algorithm’s validation is inherent in its accuracy, as segmentation was performed by an expert operator.

In the identification of stenotic lesions, we use a state-of-the-art approach based on Gaussian kernel regression. Various authors have proposed different techniques to automatically identify stenosis in coronary segmentations performed on CCTA scans; these were systematically compared in a MICCAI 2012 challenge \cite{kiricsli2013standardized}. We implemented the algorithm first proposed by Shahzad \etal \cite{shahzad2013automatic}, which achieved the highest sensitivity compared to expert observer annotations. This method was successfully implemented in several subsequent studies \cite{kiricsli2012comprehensive, fossan2018uncertainty} to investigate coronary perfusion and estimate vFFR. Compared to its original formulation, we incorporated the corrections proposed by Fossan \etal \cite{fossan2018uncertainty} for determining the parameters in Equation \ref{eq:gaussreg}, additionally introducing a regularization additive term ($\kappa$), which improved the robustness of the algorithm. In our experiments, the introduction of the constant $\kappa$ improved the model's accuracy in recognizing stenosis, compared to qualitative clinical reports. The optimized parameters $\sigma_x$ and $\sigma_r$ closely aligned with the values reported by Shahzad \etal ($\sigma_r = 0.25$ and $\sigma_x = 8$ mm) and by Fossan \etal, who described $\sigma_r$ as a function of the local radius. However, the value of $\sigma_{max}$ differed significantly in our implementation, which can be attributed to the inclusion of $\kappa$ in Equation \ref{eq:gauss_reg_k}. Unlike Fossan \textit{et al.}, we used a fixed threshold of 20\% to identify stenosis, which was extended to include points with $>$10\% narrowing. By applying the thresholding proposed by the authors (i.e., 13\% and up to 90\% narrowing of the stenosis), we observed a general overestimation of lesion severity, potentially leading to false-positive predictions.

The present study investigated the relationship between the per-vessel FAI and FFR values derived from invasive coronary angiography. The analysis revealed that, while the median FAI values were consistently higher in vessels with hemodynamically significant stenosis (FFR $\leq$ 0.80) across all three coronary arteries—RCA, LAD, LCx—, statistically significant differences were only observed for the LCx. Specifically, FAI values for stenotic LCx vessels were significantly higher compared to non-stenotic vessels (-73 vs. -78, $p=0.047$). In contrast, differences for the RCA and LAD, though directionally similar, were not statistically significant (-82 vs. -83, $p=0.655$ for RCA; -81 vs. -83.8, $p=0.463$ for LAD). These findings are partially consistent with prior work by Hoshino et al. \cite{hoshino2020peri}, who reported a significant association between FAI values measured on the LAD and FFR, suggesting that perivascular fat inflammation contributes to functional impairment in coronary flow. Similarly, Wen et al. \cite{wen_pericoronary_2021} observed significantly higher PCAT attenuation values in coronary arteries with FFR $\leq$ 0.80 (-65.6 $\pm$ 5.9 HU) compared to those with FFR $>$ 0.80 (-75.3 $\pm$ 5.4 HU,  $p<0.001$) in their vessel-level analysis across RCA, LAD, and LCx. While our study did not identify significant differences for the RCA and LAD, the trend of higher FAI values in vessels prone to ischemia closely aligns with their results. In contrast to our findings, Balcer et al. performed a 2D ROI-based analysis of FAI and did not observe significant differences in FAI values between culprit and non-culprit lesions \cite{balcer2018pericoronary}. This discrepancy may be attributed to methodological differences, particularly the use of 2D regions of interest, which may not fully capture the spatial heterogeneity of perivascular fat inflammation surrounding coronary vessels. 

At the lesion level, our analysis revealed that FAI values were significantly higher in lesions with hemodynamically significant stenosis (FFR $\leq$ 0.80) only in the RCA [-80 vs. -73 ($p = 0.003$)], while no significant differences were observed for the LAD [-82 vs. -81 ($p = 0.256$)] or LCx [-75 vs. -76 ($p = 0.883$)]. These findings contrast with those of Lihua Yu et al. \cite{yu2023radiomics}, who did not identify significant differences in FAI values between lesions causing and non-causing ischemia across all major epicardial vessels. Conversely, our results align more closely with the work of Yu et al. \cite{yu2020diagnostic} and Ma et al. \cite{ma_lesion-specific_2021}, both of whom reported significant associations between FAI and ischemic lesions. However, it is worth noting that the methodologies differ: while prior studies typically assessed all major epicardial vessels collectively, our analysis focused specifically on individual coronary arteries—RCA, LAD, and LCx. This artery-specific approach may provide a more nuanced understanding of the relationship between FAI and functional ischemia. The stronger association observed for the RCA in our study is particularly noteworthy, as it is consistent with existing evidence highlighting the prognostic significance of RCA FAI. For example, Van Diemen et al. demonstrated that only RCA FAI was independently associated with the risk of death and non-fatal myocardial infarction \cite{van_Diemen_2021_Prognostic}. Similarly, Bengs et al. reported that RCA FAI was independently predictive of major adverse coronary events (MACEs) when evaluated alongside cardiovascular risk factors, coronary computed tomography angiography (CCTA), and myocardial perfusion imaging findings \cite{bengs_quantification_2021}. These findings suggest that the RCA may have unique pathophysiological characteristics that amplify the role of perivascular fat inflammation in determining lesion significance and prognostic outcomes.

The statistical analysis conducted on PCAT features confirmed a trend previously hinted in the literature by other authors, which correlated higher mean attenuation of PCAT (i.e., tissue inflammation) with functional severity of the stenosis \cite{hoshino2020peri, yan2022pericoronary, yu2020diagnostic}. Previous studies also proposed statistical models based on univariate and multivariate analysis of various clinical parameters, including FAI, to predict FFR. We levaraged an ML-based approch for predicting potentially culprit stenosis, extending the criterion to identify ischemic lesions to other parameters. Indeed, while FFR is an established index for functional evaluation, it's been showed that ischemic lesions can be associated with $\mathrm{FFR}\geq0.8$ but WSS and $\Delta\mathrm{FFR}$ above cut-off threshold \cite{giannopoulos2018building}. This is a key novel aspect of our study, as the correlation analysis of PCAT has been limited to FFR in previous research. Our ML model, which utilizes a classic MLP backend, demonstrated good diagnostic performance by identifying culprit lesions based on FFR and WSS. Culprit lesion defined by the HRS criterion, as defined in Section \ref{sec:ml_class}, where correctly predicted only in the RCA branch (Table \ref{table:test_set_vessel}).
In general, the good performance achieved by our model in predicting functionally severe lesions based on PCAT and morphological features, which are easily quantifiable from CCTA scans, supports the hypothesis that the inflammatory status of the adipose tissue surrounding coronary vessels is associated with the functional condition of the artery itself. Inflamed adipose tissue may trigger or promote the atherosclerotic process in correspondence of the lesion site, leading to myocardial ischemia. 
The feasibility of using low-level radiomic features, such as tissue attenuation, and a few geometric characteristics—readily available once a CCTA scan is acquired—can be of great utility in cases where FFR examination cannot be performed, providing valuable insight into the hemodynamic significance of CAD for the specific patient without requiring additional examinations.
Conversely, in cases where the FFR and other functional indices are known, PCAT features can be integrated to provide deeper insights and establish a robust risk index for predicting ischemia.

\subsection{Limitations}
The size of the cohort under examination is limited compared to population studies, which typically involve hundreds of patients. The dataset of lesions under analysis included 427 cases; however, these were derived from the branches of only 72 patients. The data at hand exhibited an unbalanced representation of mild, moderate, and severe stenosis, with a higher frequency of mild cases. Despite these data biases, some trends were observed in the statistical analysis, and the ML classifier was trained with promising results.

We trained our ML model based on the results of CFD analysis conducted using an in-house simulation framework \cite{nannini2024automated} and an automatic segmentation tool \cite{nannini2024fully}. Despite the proven robustness of these two methods, we must account for possible uncertainties introduced during the computation of functional biomarkers, which are subsequently used in the PCAT-analysis pipeline.

For the patients considered in this study, clinical reports indicating stenosis grade or minimal luminal area were available for only a small subset of patients, generally providing qualitative indications. Consequently, the algorithm could not be systematically validated. Nevertheless, an overall agreement between the qualitative measurements reported by cardiologists and the quantitative analysis performed using our method was observed.

\section{Conclusion}
The inflammation status of PCAT, based on its attenuation through FAI, can serve as a biomarker to classify potentially ischemic lesions. FAI resulted on average higher in stenosis associated to functional parameters (i.e., FFR, WSS and $\Delta\mathrm{FFR}$) that indicate severe lesions, both considering the wholesome of the vessel, and focusing on the specific lesion area. PCAT features can be combined with the morphological characteristics of the lesion to inform a machine learning model for classifying potentially culprit lesions. This approach provides an indication of the functional status of the lesion without requiring additional examinations, such as invasive FFR measurement or CFD analysis.

\section*{Acknowledgment}

Funded by the European Union– Next Generation EU – NRRP M6C2 – Investment 2.1 Enhancement and strengthening of biomedical research in the NHS-Project PNRR-POC-2022-12376500.

This work was partly funded by the National Plan for NRRP Complementary Investments (PNC), established with the decree-law 6 May 2021, n. 59, converted by law n. 101 of 2021) in the call for the funding of research initiatives for technologies and innovative trajectories in the health and care sectors (Directorial Decree n. 931 of 06-06-2022) - project n. PNC0000003 - AdvaNced Technologies for Human-centrEd Medicine (project acronym: ANTHEM). This work reflects only the authors’ views and opinions, neither the Ministry for University and Research nor the European Commission can be considered responsible for them.

\ifCLASSOPTIONcaptionsoff
  \newpage
\fi

\bibliography{bibliography}{}
\bibliographystyle{ieeetr}

\end{document}